\documentclass[aps,prl,twocolumn,longbibliography]{revtex4-2} 

\usepackage{graphicx}
\usepackage{dcolumn}
\usepackage{bm}
\usepackage{times} 
\usepackage{amsmath,amssymb}
\usepackage{float}
\usepackage{color}
\usepackage{multirow}
\usepackage[normalem]{ulem}

\begin{document}
\title{Transverse thermophotovoltaics from nonreciprocal plasmon drag in metal}

\author{Dingwei He}
\affiliation{Graduate School of China Academy of Engineering Physics, Beijing 100193,
China}
\author{Gaomin Tang}
\email{gmtang@gscaep.ac.cn}
\affiliation{Graduate School of China Academy of Engineering Physics, Beijing 100193,
China}

\begin{abstract}
Transverse thermophotovoltaics has been conceptually proposed as a paradigm distinct
from conventional junction-based photovoltaics, but has so far lacked a theoretical
foundation. In this Letter, we establish a microscopic formalism of this effect in which a
transverse electric current emerges in a two-dimensional metal sheet via nonreciprocal
surface plasmon polaritons driven by near-field thermal radiation. This theoretical
formalism incorporates the electron-photon interaction by integrating electronic
transition factor governed by energy-momentum conservation, the photon flux factor
encoding the nonreciprocal surface modes, and their directional coupling.  Our approach
quantitatively confirms the plasmon-drag mechanism and reveals the role of impurity
scattering. This work provides a rigorous theoretical foundation for transverse
thermophotovoltaic devices and opens avenues for active nanoscale thermal energy
conversion. 
\end{abstract}

\maketitle

{\it Introduction.}
Transverse thermophotovoltaics, the generation of an electric current perpendicular to a
thermal gradient across two radiating objects, offers a paradigm distinct from
conventional junction-based photovoltaics. 
Recent work by Tang et al. proposed a near-field transverse thermoelectric effect by
harnessing the nonreciprocity of surface plasmon polaritons (SPPs) in a system composed of
graphene and magneto-optic medium~\cite{GT21}. In that model, the nonreciprocal dispersion
of the surface waves produces an asymmetric momentum transfer to electrons in graphene for
counterpropagating modes. This results in a directional electric current, controlled by an
external magnetic field, which emerges transversely to the thermal gradient. 
This effect fundamentally differs from traditional near-field thermophotovoltaics, which
relies on semiconductor p-n junctions to generate a longitudinal (parallel to the thermal
gradient) electric current~\cite{TPV06, TPV-review18, TPV-review21-1, TPV-review21-2,
TPV-review23}. While this work proposed a conceptual framework, a microscopic description
has been lacking.

To establish a rigorous theoretical foundation for this effect, we develop in this Letter
a microscopic formalism based on the nonequilibrium Green's function
technique~\cite{Henne08, JSW23, JSW25, Haug_Jauho}. It allows us to treat the
electron-photon interaction in a fully quantum-mechanical manner, essential for capturing
the interplay between momentum-conserving electronic transitions, nonreciprocal plasmon
dispersion, and their directional coupling through near-field radiative transfer.  A
physically transparent expression for the transverse current is obtained through a
leading-order expansion of the electron-photon interaction and electron damping. Our
theory not only confirms the earlier prediction quantitatively but also elucidates the
influence of impurity scattering that were inaccessible in previous phenomenological
descriptions. Furthermore, we propose several practical strategies to enhance the
transverse current. 

\begin{figure}
\centering
\includegraphics[width=\columnwidth]{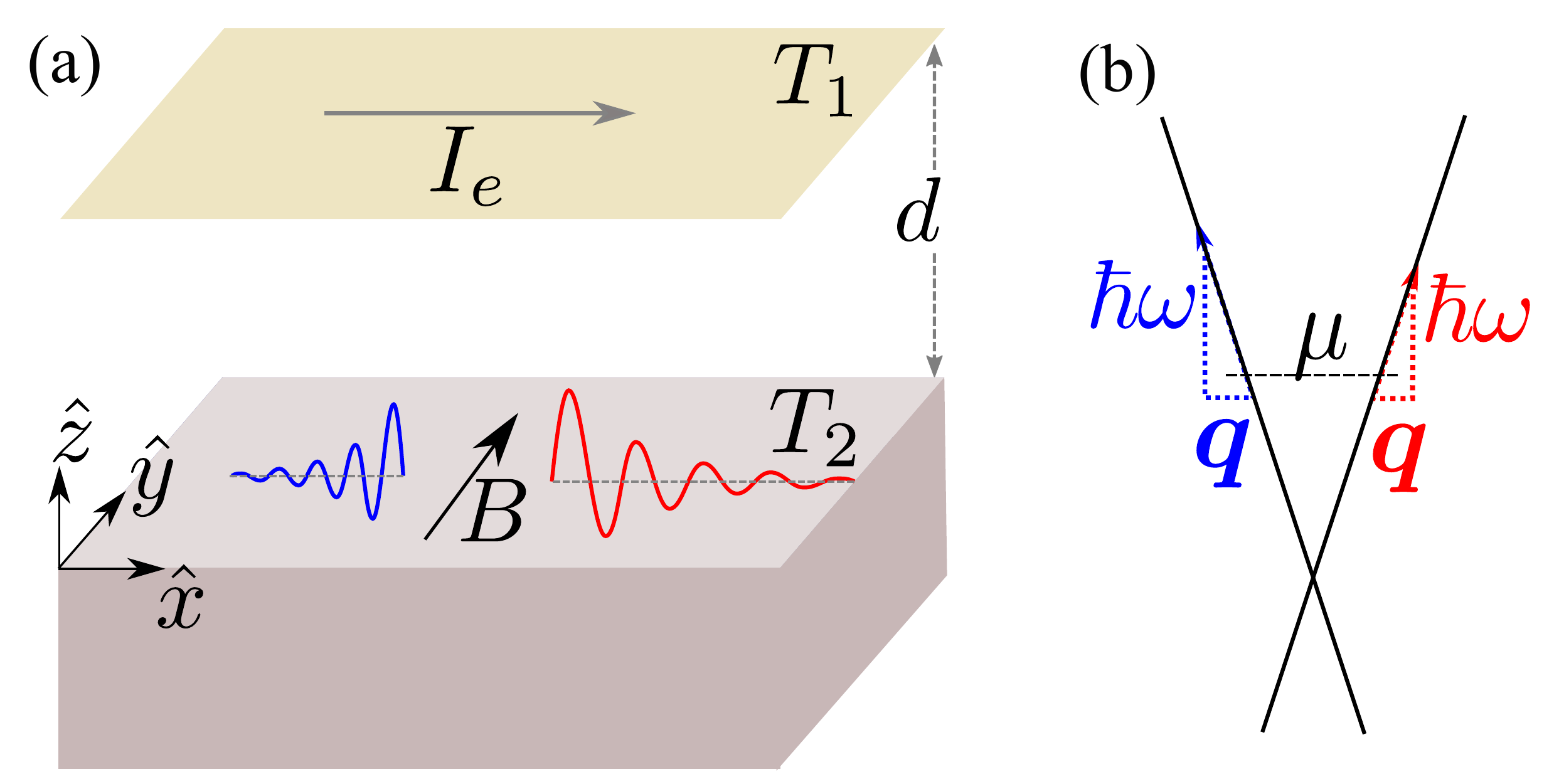} \\
\caption{(a) Schematic of the transverse thermophotovoltaic effect in a system consisting
  of a semi-infinite magneto-optic medium and a two-dimensional metal sheet at different
  temperatures, separated by a vacuum gap $d$. A magnetic field $B$ along the
  $y$ direction applied to the magneto-optic medium generates nonreciprocal surface
  plasmon polaritons (SPPs) that propagate along the $x$ direction. For a positive
  $y$-directed field, the SPP branches propagating in the $+x$ direction and $-x$
  direction are red- and blue-shifted, respectively. This spectral asymmetry creates a net
  photon-momentum transfer to electrons in the metal sheet, resulting in a transverse
  electric current powered by near-field thermal radiation. 
  (b) Nonreciprocal electronic transition in a single-layer graphene with linear
  dispersion by absorption of a near-field photon with large in-plane momentum
  $\hbar\bm{q}$. The photon transfers energy $\hbar\omega$ and momentum $\hbar\bm{q}$ to
  an electron, scattering it from state $|\bm{k}\rangle$ to $|\bm{k}+\bm{q}\rangle$. Since
  the incident photon flux originates from the nonreciprocal SPPs, the transition
  probability differs for $+q_x$ and $-q_x$. This imbalance yields a net momentum transfer
  along the $x$-direction, underlying the generation of the transverse current shown in
  panel (a). }
\label{fig1}
\end{figure}

{\it Theoretical formalism.}
We model the transverse thermophotovoltaic effect in a system comprising a semi-infinite
magneto-optic medium which fills the space of $z<0$ and a two-dimensional metal sheet
separated by a vacuum gap of distance $d$, as sketched in Fig.~\ref{fig1}(a). 
The metal sheet and the magneto-optic medium are at different temperatures, $T_1$ and
$T_2$, respectively.
The magneto-optic medium, taken to be InSb, is subjected to an external magnetic field
along the $y$ direction. Its dielectric response at frequency $\omega$ is described by the
gyrotropic tensor~\cite{InSb_76}
\begin{equation} \label{epsilon_MO}
  \epsilon_{\rm MO}(\omega) = 
  \begin{bmatrix}
    \epsilon_d & 0 & i\epsilon_a \\
    0 & \epsilon_p & 0 \\
    -i\epsilon_a & 0 & \epsilon_d
  \end{bmatrix} .
\end{equation}
This gyrotropy breaks time-reversal symmetry and supports nonreciprocal SPPs that
propagate along the $x$ direction. 
An alternative route to nonreciprocity exploits materials with intrinsic magneto-optical
activity, such as magnetic Weyl semimetals. These systems host nonreciprocal surface
polaritons without an external field, owing to their inherent anomalous Hall
conductivity~\cite{WSM_SPP16, Kotov18, WSM_AHE_20nc, GT_WSM}.

The total Hamiltonian is decomposed as $H_{\rm tot} = H_{\rm ph} + H_1 + H_{1, {\rm ph}}$.
The electromagnetic environment, including the magneto-optic medium and the free space, is
described by $H_{\rm ph}$. The metal sheet is described by the electronic Hamiltonian
$H_1$ and its coupling to the electromagnetic field by $H_{1, {\rm ph}}$. The electric
current induced by nonreciprocal plasmon drag is computed within the nonequilibrium
Green's function formalism~\cite{Haug_Jauho}.
We neglect Zeeman splitting induced by the magnetic field, and the spin and
valley degeneracies are incorporated by an overall factor of $4$ in the electric current
expression. Expanding to the second order in the electron-photon interaction strength and
to the first order in electron damping parameter $\eta$ in the metal, the steady state
electric current along $x$ direction is expressed as [see End Matter for details]
\begin{equation} \label{Ie}
  I_e = -\frac{4 e_0}{\eta} \int d X \ \Delta_x(\bm{k},\bm{q}) L(\bm{k}, \bm{q}, \omega)
  M(\bm{k}, \bm{q}, \omega) N_{21}(\omega) ,
\end{equation}
where $e_0$ is the elementary charge and the integration measure is 
\begin{equation*}
  \int dX = \int \frac{d^2\bm{k}}{(2\pi)^2} \int_{q>\omega/c}
  \frac{d^2\bm{q}}{(2\pi)^2} \int_0^\infty \frac{d\omega}{2\pi} .
\end{equation*}
The integration runs over the electron wavevector $\bm{k}$, the in-plane photon wavevector
$\bm{q}\equiv (q_x, q_y)$, and photon frequency $\omega$. 
The restriction $q \equiv |\bm{q}|>\omega / c$ with $c$ the speed of light focuses the
discussion on the near-field region, where evanescent waves dominate. 
The velocity difference $\bm{\Delta}(\bm{k},\bm{q}) = \bm{v}(\bm{k} + \bm{q}) 
- \bm{v}(\bm{k})$
with $\bm{v}(\bm{k})$ the electron group velocity enters as an electron scatter from
state $|\bm{k}\rangle$ to state $|\bm{k} + \bm{q}\rangle$ by absorbing a photon of
momentum $\hbar\bm{q}$. 
The electronic transition factor
\begin{equation} \label{L}
  L(\bm{k},\bm{q}, \omega) = {\rm Im} \left[ \frac{f(\varepsilon_{\bm{k}}-i
  \eta)-f(\varepsilon_{\bm{k}+\bm{q}}+i \eta)}{\hbar\omega +
  \varepsilon_{\bm{k}}-\varepsilon_{\bm{k}+\bm{q}} -2i \eta} \right] ,
\end{equation}
where $\varepsilon_{\bm{k}}$ is the electronic dispersion, and 
\begin{equation} \label{Fermi-Dirac}
  f(E) = \frac{1}{\exp[(E-\mu)/(k_B T_1)]+1}
\end{equation}
is the Fermi-Dirac distribution with chemical potential $\mu$, gives the energy- and
momentum-resolved transition probability density for photon absorption. 
The denominator of $L$ ensures that the transition is most likely when the photon energy
$\hbar\omega$ matches the energy difference between the final electronic state $|\bm{k} +
\bm{q}\rangle$ and the initial one $|\bm{k} \rangle$ [See Fig.~\ref{fig1}(b)]. 
Since the electronic transition factor depends on the band dispersion, which also dictates
the electron group velocity, a microscopic framework beyond the conventional fluctuational
electrodynamics is essential for modeling the electric-current generation.
The photon interaction strength
\begin{equation} \label{M}
  M(\bm{k},\bm{q}, \omega) = \frac{e_0^2 \hbar |\gamma_0|}{\epsilon_0 \omega^2}
  \Phi(\bm{q}, \omega) \big[ \bm{v}(\bm{k},\bm{q}) \cdot \hat{\bm{q}} \big]^2 ,
\end{equation}
with $\epsilon_0$ the vacuum permittivity and $\gamma_0 = \sqrt{(\omega/c)^2-q^2}$ the
out-of-plane decay constant, quantifies the coupling between the evanescent
electromagnetic field and the electrons in the metal. 
The symmetrized velocity is given by $\bm{v}(\bm{k},\bm{q}) = [ \bm{v}(\bm{k}) 
+ \bm{v}(\bm{k} + \bm{q})] /
2$.
The photon flux factor
\begin{equation}
  \Phi(\bm{q}, \omega) = \frac{{\rm Im}\big( r_2^p \big) |t_1^p|^2 e^{-2 |\gamma_0|
  d}}{|1 - r_1^p r_2^p e^{-2 |\gamma_0| d}|^2} 
\end{equation}
describes the evanescent wave that reaches the metal sheet.
Here $r_2^p$ is the Fresnel reflection coefficient for $p$-polarization waves incident from 
vacuum onto the magneto-optic interface, which is the source of the nonreciprocity. Its
calculation detail is provided in the Supporting Information of Ref.~\cite{GT_WSM}. 
The reflection and transmission coefficients at the vacuum-metal interface are determined
by the sheet conductivity $\sigma(\omega)$:
\begin{equation} \label{r1p}
  r^p_1(q, \omega) = 1- t^p_1(q, \omega) 
  = \frac{\gamma_0\sigma( \omega)}{2\epsilon_0\omega + \gamma_0\sigma( \omega)} .
\end{equation}
The directional factor $\bm{v}(\bm{k},\bm{q}) \cdot \hat{\bm{q}}$ reflects
that only the component of the electron velocity parallel to the photon momentum
$\hat{\bm{q}}$ couples to the $p$-polarized electric field. In our geometry,
nonreciprocity occurs along the $x$ axis, and this selection rule ensures that the induced
electric current flows in the same direction. The thermodynamic drive for the current
generation is provided by the Bose–Einstein distribution difference between the two
bodies:
\begin{equation} \label{N21}
  N_{21}(\omega) = \frac{1}{\exp(\hbar\omega/k_B T_2)-1} -
  \frac{1}{\exp(\hbar\omega/k_B T_1)-1} .
\end{equation}
The sign of $N_{21}(\omega)$ determines the direction of the heat flux and hence the
momentum transferred to the electrons. Consequently, reversing the temperature gradient
reverses the current direction.

{\it Numerical results.}
For the numerical calculations, 
the components of the dielectric tensor in Eq.~\eqref{epsilon_MO} are given by
\begin{align*}
  \epsilon_d &= \epsilon_{\infty} \left\{ 1+ \frac{\omega_L^2 -\omega_T^2}{\omega_T^2
  -\omega^2 -i\Gamma\omega} + \frac{\omega_p^2(\omega+i\gamma)}{\omega\big[
\omega_c^2-(\omega+i\gamma)^2 \big]} \right\}, \\
  \epsilon_p &= \epsilon_{\infty} \left[ 1+ \frac{\omega_L^2 -\omega_T^2}{\omega_T^2
  -\omega^2 -i\Gamma\omega} - \frac{\omega_p^2}{\omega(\omega+i\gamma)} \right], \\
  \epsilon_a &= \frac{\epsilon_{\infty}\omega_p^2\omega_c}{\omega\big[ (\omega+i\gamma)^2
  -\omega_c^2 \big]} .
\end{align*}
The parameters of InSb are taken from Ref.~\cite{InSb_76} with
the high-frequency dielectric constant $\epsilon_\infty = 15.7$, 
the longitudinal optical phonon frequency $\omega_L = 3.62\times 10^{13}\,{\rm rad/s}$, 
the transverse optical phonon frequency $\omega_T = 3.39\times 10^{13}\,{\rm rad/s}$, 
the phonon damping constant $\Gamma = 5.65\times 10^{11}\,{\rm rad/s}$, 
the free-carrier damping constant $\gamma = 3.39\times 10^{12}\,{\rm rad/s}$, 
the plasma frequency $\omega_p = 3.14\times 10^{13}\,{\rm rad/s}$, and
the cyclotron frequency $\omega_c = 8.02\times 10^{12}\,{\rm rad/s}$ for $B=1\,$T.

The metal sheet is modeled as single-layer graphene with linear dispersion
$\varepsilon_{\bm{k}} = \hbar v_F |\bm{k}|$ and Fermi velocity $v_F = 10^6\,$m/s. The
chemical potential is fixed at $\mu = 0.15\,$eV. The corresponding Fermi wavevector is
$k_F=2.28\times 10^8\,$m$^{-1}$, and the electron density which includes spin and valley
degeneracies is given by $n_s = k_F^2/\pi = 1.65\times 10^{16}\,$m$^{-2}$. Since the
chemical potential is much larger than the SPP energies considered, interband transitions
are negligible in the frequency range of interest. Consequently, the optical conductivity
is well described by the Drude form 
$\sigma(\omega)= e_0^2 \mu /[\hbar \pi (2\eta -i\hbar\omega)]$.

Before presenting the numerical results, we estimate the vacuum gap required for efficient
electron–photon coupling under energy and momentum conservation. Since the dominant
electronic states contributing to transport reside near the Fermi level, we take the
initial electron momentum as the Fermi momentum $k_F$. Absorption of a surface plasmon
with in-plane wavevector $q$ and energy $\hbar\omega$ imposes the condition 
$q=\omega /v_F$. For SPP energy $\hbar\omega$ at $30\,$meV, the evanescent decay length in
vacuum is thus $l \sim 1/q \approx 22\,$nm. To ensure substantial photon tunneling across
the gap, the vacuum gap $d$ should be smaller than this length scale.

\begin{figure}
\centering
\includegraphics[width=\columnwidth]{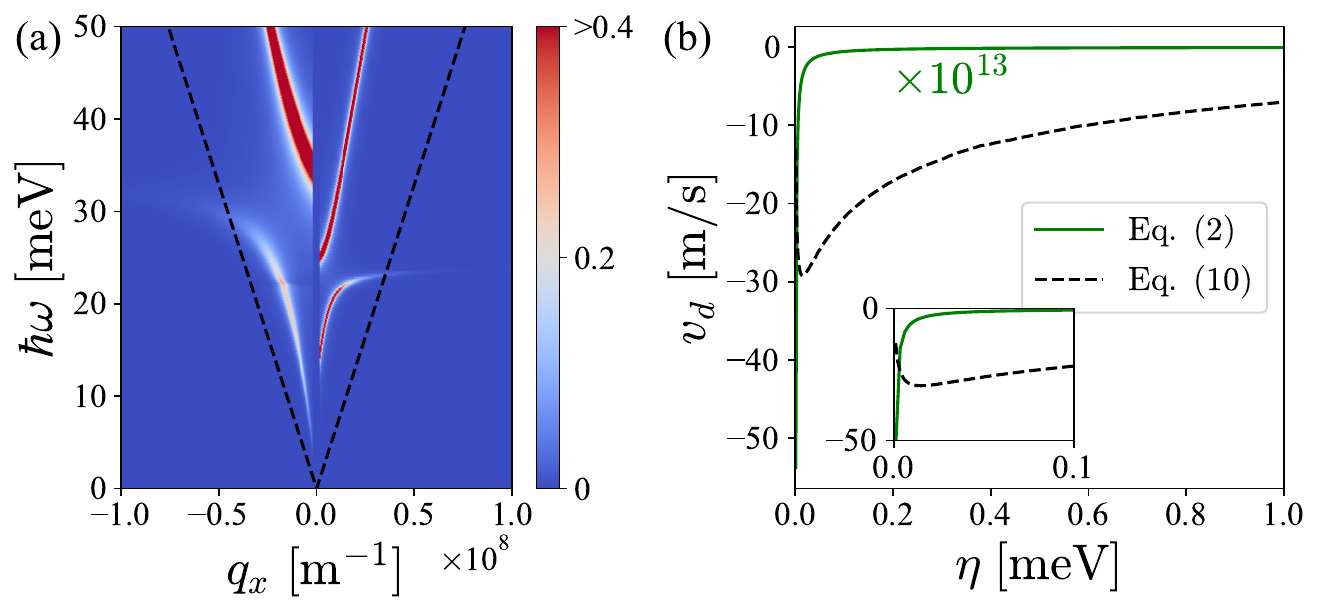} \\
\caption{(a) Nonreciprocal photon flux factor $\Phi(q_x, \omega)$ plotted as a function of
  in-plane wavevector $q_x$ and photon energy $\hbar\omega$ for $q_y=0$ with magnetic
  field $B=4\,$T, electron damping $\eta =0.5\,$meV, and vacuum gap $d=2\,$nm.
  Photon absorption is strongly peaked along the dashed lines, which correspond to the
  single-particle excitations in graphene, given by $\omega = |q_x| v_F$ (dashed lines).
  (b) The effective drift velocity versus $\eta$. The temperatures of the magneto-optic
  medium and graphene are set to $T_2=400\,$K and $T_1=300\,$K, respectively. For visual
  clarity, the curve from Eq.~\eqref{Ie} is scaled by a factor of $10^{13}$.  } 
\label{fig2}
\end{figure}

In Fig.~\ref{fig2}(a), we show the nonreciprocal photon flux factor $\Phi(\bm{q},\omega)$
as a function of in-plane wavevector $q_x$ (with $q_y=0$) and SPP energy $\hbar\omega$
with magnetic field $B=4\,$T, electronic damping $\eta = 0.5\,$meV, and vacuum gap
$d=2\,$nm. The dashed lines correspond to the single-particle excitation condition in
graphene, $\omega = |q_x| v_F$, along which photon absorption is the strongest. From
Eq.~\eqref{L}, the dominant contribution to the electron transport comes from the region
in $(\bm{q}, \omega)$ space where the dispersion of the SPP intersects the single-particle
excitation spectrum. Because the electron spectrum is sharply peaked along $\omega =
|\bm{q}| v_F$, only those SPP modes that lie close to this line efficiently couple to the
electrons and contribute to current generation. Figure~\ref{fig2}(a) shows that the SPP
branch propagating in the $-x$ direction ($q_x<0$) possesses a larger spectral weight near
the single-particle excitation line than the branch propagating in the $+x$ direction
($q_x>0$). This difference originates from the nonreciprocal SPP shift, sketched in
Fig.~\ref{fig1}. As a result, plasmons with $q_x <0$ are absorbed more efficiently,
producing a net electron drift along the $-x$ direction. 

\begin{figure}
\centering
\includegraphics[width=\columnwidth]{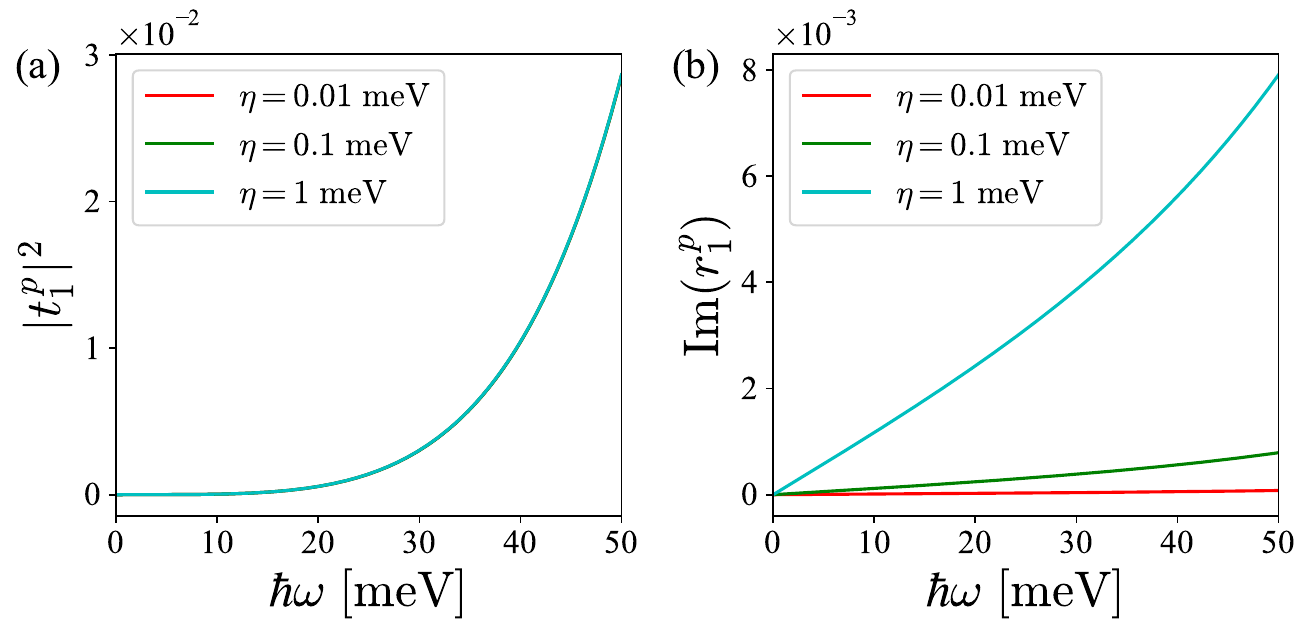} \\
\caption{Fresnel coefficients for $p$-polarized waves incident on graphene as a function of
  $\hbar\omega$ for $q=4\times 10^7\,$m$^{-1}$ and various electron damping $\eta$. 
  The squared transmission coefficient $|t_1^p|^2$ exhibits negligible variation with $\eta$
  on the scale shown. The imaginary part of the reflection coefficient, ${\rm Im}(r_1^p)$,
  increases monotonically with $\eta$. }
\label{fig3}
\end{figure}

To have an intuitive measure, an effective drift velocity $v_d$ is introduced via 
\begin{equation} \label{Ie_vd}
  I_e = -n_s e_0 v_d .
\end{equation}
Figure~\ref{fig2}(b) displays $v_d$ as a function of the electron damping parameter
$\eta$. The negative sign of $v_d$ confirms the electron flow direction inferred from
Fig.~\ref{fig2}(a).
The drift velocity magnitude decreases with increasing $\eta$ due to enhanced
momentum dissipation. The photon flux factor, governed by $|t_1^p|$, remains insensitive
to $\eta$ and thus does not counter this trend [See Fig.~\ref{fig3}(a)].
The drift velocity is small because only a narrow region of the plasmon spectrum satisfies
the kinematic matching for electronic transitions and contributes to the current.
The drift velocity is approximately $-0.05\,$pm/s for $\eta=0.1\,$meV. The
corresponding current density is approximately $0.14\,$fA/m.

We now provide numerical evidence demonstrating why a microscopic treatment is essential.
The total photon momentum flux transferred to the metal sheet can be written in a
Landauer-like form as~\cite{SM}
\begin{equation} \label{Px}
  P_x = \int_0^{\infty} \frac{d\omega}{2\pi} \int_{q>
  \omega/c}\frac{d^2\bm{q}}{(2\pi)^2} \hbar q_x \xi(\bm{q} ,\omega) N_{21}(\omega) ,
\end{equation}
with the photon transmission coefficient 
\begin{equation}
  \xi(\bm{q},\omega) = \frac{4 \ {\rm Im}\big(r_2^p \big) {\rm Im}\big( r_1^p \big)
  e^{-2 |\gamma_0| d}}{|1 - r_1^p r_2^p e^{-2 |\gamma_0| d}|^2}.
\end{equation}
This expression has been employed to calculate the lateral force induced by the
nonreciprocal surface mode~\cite{lateral_17, lateral_21, GT24-2}; a microscopic derivation
is provided in the Supplemental Material~\cite{SM}. 
Momentum conservation dictates that photons directly transfer momentum to the electrons.
This momentum is subsequently dissipated to the lattice through scattering, which
establishes the relaxation mechanism necessary for a steady electric current to develop.
Thus, we might employ the simple force-balance relation $P_x = m^* n_s v_d /\tau$ which
assumes a constant scattering time with $\tau =\hbar/(2\eta)$. 
However, the momentum- and energy-conservation conditions shown in Eq.~\eqref{L} and
Fig.~\ref{fig2}(a) imply that only a narrow region of the plasmon spectrum satisfies the
kinematic matching for electronic transitions and thus contributes to the current. A
simple force-balance equation that ignores this selective coupling is therefore
inadequate.

Moreover, the photon flux factor $\Phi$ entering Eq.~\eqref{Ie} and the 
photon transmission coefficient $\xi$ in Eq.~\eqref{Px} have fundamentally different
physical meaning: $\Phi$ quantifies the flux of photons that can actually contribute to the
electronic transitions and drive an electron flow, while $\xi$ describes the photon 
dissipation in the metal. This difference can also be reflected in the fact that, in the
limit of vanishing electronic damping ($\eta \rightarrow 0$), the
photon flux factor $\Phi$ remains finite, whereas the photon transmission coefficient
$\xi$ goes to zero [See Fig.~\ref{fig3}].
The electron-photon interaction, encoded in the Fock self-energy [see End Matter],
also renormalizes essential electronic parameters, including the scattering rate, the
carrier density, and the effective mass. Importantly, this self-energy depends on the
temperature profile and can vary over the energy range relevant to transport.

For a quantitative comparison, the dashed curve in Fig.~\ref{fig2}(b) shows the drift
velocity obtained from the simple force-balance relation $P_x = m^* n_s v_d /\tau$
where the effective mass $m^*$ is taken as $m^* =\hbar k_F / v_F \approx 0.026m_e$
with $m_e$ the free-electron mass. The resulting values overestimate the microscopic
result by approximately $13$ orders of magnitude, demonstrating that a force-balance
picture fails utterly to describe the coupled electron-photon dynamics. 
Moreover, the magnitude of the drift velocity calculated using this approch exhibits a
nonmonotonic dependence on the scattering rate $\eta$: it first increases and then
decreases. This is due to the competition between photon transmission and momentum
relaxation. At very small $\eta$, the photon transmission is low, which limits the photon
momentum flux $P_x$, as indicated by the behavior of ${\rm Im}(r_1^p)$ [See
Fig.~\ref{fig3}(b)]. 
As $\eta$ increases, $P_x$ initially rises. However, beyond a certain value, the explicit
$1/\eta$ factor representing increased momentum dissipation dominates and suppresses the
electron drift.
These drastic discrepancies underscore the necessity of a microscopic theory that properly
incorporates electron-photon interaction in transverse thermophotovoltaic effects.

Photons that do not contribute to the current are absorbed via Drude absorption: the
oscillating electric field transfers energy to the electron gas by driving collective
oscillations. The excited electrons then relax through electron-phonon coupling,
converting energy to heat and transferring photon momentum to the lattice. This relaxation
process is characterized by $\eta$; indeed, both the radiative heat flux and total
photon momentum flux vanish as $\eta \rightarrow 0$. Crucially, this mechanical force
does not affect the current in Eq.~\eqref{Ie}, which describes electron motion relative to
the graphene lattice.

{\it Discussion and conclusion.}
Regarding experimental feasibility, the predicted transverse current may be too small for
direct detection with current experimental technique. To raise the signal to a measurable
level, two principal routes can be pursued: enhancing the photon flux factor and
optimizing the electronic transition.

By placing a second magneto-optic medium under identical magnetic field and temperature
conditions on the opposite side of the metal sheet, the resulting Fabry-Perot resonance
between the two media can substantially enhance the photon flux factor.

A strategy to simultaneously realize both enhancements is to pattern periodic gratings on
the metal sheet. First, the grating periodicity folds the plasmon dispersion, opening
photonic band gaps that suppress SPP excitation for one propagation direction while
allowing it in the opposite direction, thereby increasing the momentum-transfer asymmetry.
Second, the grating grooves act as plasmonic resonators that trap and locally enhance the
electromagnetic field, increasing the photon flux factor. Through these two synergistic
mechanisms, the transverse electric current can be increased.

An alternative approach is to replace the metal sheet with an intrinsic semiconductor or a
gated two-dimensional electron gas. The finite band gap (or the gate-tunable excitation
threshold) provides a sharp optical absorption edge. Because the nonreciprocal SPP
branches are shifted in frequency, one branch can lie above the absorption edge while the
other falls below it. Consequently, the semiconductor strongly absorbs SPPs propagating in
one direction while blocking those in the opposite direction, dramatically enhancing the
directional asymmetry of the photon-momentum transfer. 
Thus, using a semiconductor as the current generator can alleviate heat dissipation by
improving thermoelectric conversion efficiency through relaxed kinematic matching. 
This lifts the constraint present in metals where only a narrow intersection region
contributes and contributions from opposite propagation directions largely cancel.

To assess the energy conversion efficiency and output power of this heat engine would
require attaching two metallic leads to the graphene sheet. Under operating conditions,
the induced voltage modifies the electronic states, making the local thermal equilibrium
assumption fail. Consequently, a framework that self-consistently accounts for the
nonequilibrium carrier distribution would be necessary.
The same nonequilibrium considerations apply to a signal-to-noise analysis, which would
require including electron-phonon scattering and other noise sources.

The transverse thermophotovoltaic effect reported in this work differs from conventional
photon drag~\cite{photon_drag88, photon_drag19, photon_drag24, photon_drag25} in three
aspects: (i) the photons are thermal near-field photons with large in-plane momenta ($q\gg
\omega/c$), not coherent laser photons; (ii) the electric current is driven by a
temperature gradient, not an external optical beam; (iii) the directionality comes from
intrinsic SPP nonreciprocity, not sample asymmetry or oblique incidence. 

To conclude, we have demonstrated the generation of a transverse electric current driven
by near-field thermal radiation via nonreciprocal surface plasmon polaritons, based on a
microscopic theory that accounts for electron-photon interactions. 
The derived electric current expression, Eq.~\eqref{Ie}, inherently incorporates kinematic
matching and momentum-resolved photon flux, providing a rigorous theoretical foundation
for this phenomena. This work establishes a design principle to harness the nonreciprocal
surface modes for ``transverse thermophotovoltaics" with potential applications in
nanoscale energy conversion and active radiative thermal management. 


{\it Note added.} During the preparation of this manuscript, we became aware of a relevant
study~\cite{photon_Nernst25}.

{\it Acknowledgments.}
D.H. and G.T. acknowledge the support from National Natural Science Foundation of China
(Grants No. 12374048 and No. 12088101).


\begin{thebibliography}{29}%
\makeatletter
\providecommand \@ifxundefined [1]{%
 \@ifx{#1\undefined}
}%
\providecommand \@ifnum [1]{%
 \ifnum #1\expandafter \@firstoftwo
 \else \expandafter \@secondoftwo
 \fi
}%
\providecommand \@ifx [1]{%
 \ifx #1\expandafter \@firstoftwo
 \else \expandafter \@secondoftwo
 \fi
}%
\providecommand \natexlab [1]{#1}%
\providecommand \enquote  [1]{``#1''}%
\providecommand \bibnamefont  [1]{#1}%
\providecommand \bibfnamefont [1]{#1}%
\providecommand \citenamefont [1]{#1}%
\providecommand \href@noop [0]{\@secondoftwo}%
\providecommand \href [0]{\begingroup \@sanitize@url \@href}%
\providecommand \@href[1]{\@@startlink{#1}\@@href}%
\providecommand \@@href[1]{\endgroup#1\@@endlink}%
\providecommand \@sanitize@url [0]{\catcode `\\12\catcode `\$12\catcode
  `\&12\catcode `\#12\catcode `\^12\catcode `\_12\catcode `\%12\relax}%
\providecommand \@@startlink[1]{}%
\providecommand \@@endlink[0]{}%
\providecommand \url  [0]{\begingroup\@sanitize@url \@url }%
\providecommand \@url [1]{\endgroup\@href {#1}{\urlprefix }}%
\providecommand \urlprefix  [0]{URL }%
\providecommand \Eprint [0]{\href }%
\providecommand \doibase [0]{https://doi.org/}%
\providecommand \selectlanguage [0]{\@gobble}%
\providecommand \bibinfo  [0]{\@secondoftwo}%
\providecommand \bibfield  [0]{\@secondoftwo}%
\providecommand \translation [1]{[#1]}%
\providecommand \BibitemOpen [0]{}%
\providecommand \bibitemStop [0]{}%
\providecommand \bibitemNoStop [0]{.\EOS\space}%
\providecommand \EOS [0]{\spacefactor3000\relax}%
\providecommand \BibitemShut  [1]{\csname bibitem#1\endcsname}%
\let\auto@bib@innerbib\@empty
\bibitem [{\citenamefont {Tang}\ \emph
  {et~al.}(2021{\natexlab{a}})\citenamefont {Tang}, \citenamefont {Zhang},
  \citenamefont {Zhang}, \citenamefont {Chen},\ and\ \citenamefont
  {Chan}}]{GT21}%
  \BibitemOpen
  \bibfield  {author} {\bibinfo {author} {\bibfnamefont {G.}~\bibnamefont
  {Tang}}, \bibinfo {author} {\bibfnamefont {L.}~\bibnamefont {Zhang}},
  \bibinfo {author} {\bibfnamefont {Y.}~\bibnamefont {Zhang}}, \bibinfo
  {author} {\bibfnamefont {J.}~\bibnamefont {Chen}},\ and\ \bibinfo {author}
  {\bibfnamefont {C.~T.}\ \bibnamefont {Chan}},\ }\bibfield  {title} {\bibinfo
  {title} {Near-field energy transfer between graphene and magneto-optic
  media},\ }\href {https://doi.org/10.1103/PhysRevLett.127.247401} {\bibfield
  {journal} {\bibinfo  {journal} {Phys. Rev. Lett.}\ }\textbf {\bibinfo
  {volume} {127}},\ \bibinfo {pages} {247401} (\bibinfo {year}
  {2021}{\natexlab{a}})}\BibitemShut {NoStop}%
\bibitem [{\citenamefont {Laroche}\ \emph {et~al.}(2006)\citenamefont
  {Laroche}, \citenamefont {Carminati},\ and\ \citenamefont {Greffet}}]{TPV06}%
  \BibitemOpen
  \bibfield  {author} {\bibinfo {author} {\bibfnamefont {M.}~\bibnamefont
  {Laroche}}, \bibinfo {author} {\bibfnamefont {R.}~\bibnamefont {Carminati}},\
  and\ \bibinfo {author} {\bibfnamefont {J.-J.}\ \bibnamefont {Greffet}},\
  }\bibfield  {title} {\bibinfo {title} {Near-field thermophotovoltaic energy
  conversion},\ }\href {https://doi.org/10.1063/1.2234560} {\bibfield
  {journal} {\bibinfo  {journal} {J. Appl. Phys.}\ }\textbf {\bibinfo {volume}
  {100}},\ \bibinfo {pages} {063704} (\bibinfo {year} {2006})}\BibitemShut
  {NoStop}%
\bibitem [{\citenamefont {Tervo}\ \emph {et~al.}(2018)\citenamefont {Tervo},
  \citenamefont {Bagherisereshki},\ and\ \citenamefont {Zhang}}]{TPV-review18}%
  \BibitemOpen
  \bibfield  {author} {\bibinfo {author} {\bibfnamefont {E.}~\bibnamefont
  {Tervo}}, \bibinfo {author} {\bibfnamefont {E.}~\bibnamefont
  {Bagherisereshki}},\ and\ \bibinfo {author} {\bibfnamefont {Z.}~\bibnamefont
  {Zhang}},\ }\bibfield  {title} {\bibinfo {title} {Near-field radiative
  thermoelectric energy converters: a review},\ }\href@noop {} {\bibfield
  {journal} {\bibinfo  {journal} {Front. Energy}\ }\textbf {\bibinfo {volume}
  {12}},\ \bibinfo {pages} {5} (\bibinfo {year} {2018})}\BibitemShut {NoStop}%
\bibitem [{\citenamefont {Callahan}\ \emph {et~al.}(2021)\citenamefont
  {Callahan}, \citenamefont {Feng}, \citenamefont {Zhang}, \citenamefont
  {Toberer}, \citenamefont {Ferguson},\ and\ \citenamefont
  {Tervo}}]{TPV-review21-1}%
  \BibitemOpen
  \bibfield  {author} {\bibinfo {author} {\bibfnamefont {W.~A.}\ \bibnamefont
  {Callahan}}, \bibinfo {author} {\bibfnamefont {D.}~\bibnamefont {Feng}},
  \bibinfo {author} {\bibfnamefont {Z.~M.}\ \bibnamefont {Zhang}}, \bibinfo
  {author} {\bibfnamefont {E.~S.}\ \bibnamefont {Toberer}}, \bibinfo {author}
  {\bibfnamefont {A.~J.}\ \bibnamefont {Ferguson}},\ and\ \bibinfo {author}
  {\bibfnamefont {E.~J.}\ \bibnamefont {Tervo}},\ }\bibfield  {title} {\bibinfo
  {title} {Coupled charge and radiation transport processes in
  thermophotovoltaic and thermoradiative cells},\ }\href
  {https://doi.org/10.1103/PhysRevApplied.15.054035} {\bibfield  {journal}
  {\bibinfo  {journal} {Phys. Rev. Appl.}\ }\textbf {\bibinfo {volume} {15}},\
  \bibinfo {pages} {054035} (\bibinfo {year} {2021})}\BibitemShut {NoStop}%
\bibitem [{\citenamefont {Papadakis}\ \emph {et~al.}(2021)\citenamefont
  {Papadakis}, \citenamefont {Orenstein}, \citenamefont {Yablonovitch},\ and\
  \citenamefont {Fan}}]{TPV-review21-2}%
  \BibitemOpen
  \bibfield  {author} {\bibinfo {author} {\bibfnamefont {G.~T.}\ \bibnamefont
  {Papadakis}}, \bibinfo {author} {\bibfnamefont {M.}~\bibnamefont
  {Orenstein}}, \bibinfo {author} {\bibfnamefont {E.}~\bibnamefont
  {Yablonovitch}},\ and\ \bibinfo {author} {\bibfnamefont {S.}~\bibnamefont
  {Fan}},\ }\bibfield  {title} {\bibinfo {title} {Thermodynamics of light
  management in near-field thermophotovoltaics},\ }\href
  {https://doi.org/10.1103/PhysRevApplied.16.064063} {\bibfield  {journal}
  {\bibinfo  {journal} {Phys. Rev. Appl.}\ }\textbf {\bibinfo {volume} {16}},\
  \bibinfo {pages} {064063} (\bibinfo {year} {2021})}\BibitemShut {NoStop}%
\bibitem [{\citenamefont {Mittapally}\ \emph {et~al.}(2023)\citenamefont
  {Mittapally}, \citenamefont {Majumder}, \citenamefont {Reddy},\ and\
  \citenamefont {Meyhofer}}]{TPV-review23}%
  \BibitemOpen
  \bibfield  {author} {\bibinfo {author} {\bibfnamefont {R.}~\bibnamefont
  {Mittapally}}, \bibinfo {author} {\bibfnamefont {A.}~\bibnamefont
  {Majumder}}, \bibinfo {author} {\bibfnamefont {P.}~\bibnamefont {Reddy}},\
  and\ \bibinfo {author} {\bibfnamefont {E.}~\bibnamefont {Meyhofer}},\
  }\bibfield  {title} {\bibinfo {title} {Near-field thermophotovoltaic energy
  conversion: Progress and opportunities},\ }\href
  {https://doi.org/10.1103/PhysRevApplied.19.037002} {\bibfield  {journal}
  {\bibinfo  {journal} {Phys. Rev. Appl.}\ }\textbf {\bibinfo {volume} {19}},\
  \bibinfo {pages} {037002} (\bibinfo {year} {2023})}\BibitemShut {NoStop}%
\bibitem [{\citenamefont {Richter}\ \emph {et~al.}(2008)\citenamefont
  {Richter}, \citenamefont {Florian},\ and\ \citenamefont
  {Henneberger}}]{Henne08}%
  \BibitemOpen
  \bibfield  {author} {\bibinfo {author} {\bibfnamefont {F.}~\bibnamefont
  {Richter}}, \bibinfo {author} {\bibfnamefont {M.}~\bibnamefont {Florian}},\
  and\ \bibinfo {author} {\bibfnamefont {K.}~\bibnamefont {Henneberger}},\
  }\bibfield  {title} {\bibinfo {title} {Generalized radiation law for excited
  media in a nonequilibrium steady state},\ }\href
  {https://doi.org/10.1103/PhysRevB.78.205114} {\bibfield  {journal} {\bibinfo
  {journal} {Phys. Rev. B}\ }\textbf {\bibinfo {volume} {78}},\ \bibinfo
  {pages} {205114} (\bibinfo {year} {2008})}\BibitemShut {NoStop}%
\bibitem [{\citenamefont {Wang}\ \emph {et~al.}(2023)\citenamefont {Wang},
  \citenamefont {Peng}, \citenamefont {Zhang}, \citenamefont {Zhang},\ and\
  \citenamefont {Zhu}}]{JSW23}%
  \BibitemOpen
  \bibfield  {author} {\bibinfo {author} {\bibfnamefont {J.-S.}\ \bibnamefont
  {Wang}}, \bibinfo {author} {\bibfnamefont {J.}~\bibnamefont {Peng}}, \bibinfo
  {author} {\bibfnamefont {Z.-Q.}\ \bibnamefont {Zhang}}, \bibinfo {author}
  {\bibfnamefont {Y.-M.}\ \bibnamefont {Zhang}},\ and\ \bibinfo {author}
  {\bibfnamefont {T.}~\bibnamefont {Zhu}},\ }\bibfield  {title} {\bibinfo
  {title} {Transport in electron-photon systems},\ }\href@noop {} {\bibfield
  {journal} {\bibinfo  {journal} {Front. Phys.}\ }\textbf {\bibinfo {volume}
  {18}},\ \bibinfo {pages} {43602} (\bibinfo {year} {2023})}\BibitemShut
  {NoStop}%
\bibitem [{\citenamefont {Wang}(2025)}]{JSW25}%
  \BibitemOpen
  \bibfield  {author} {\bibinfo {author} {\bibfnamefont {J.-S.}\ \bibnamefont
  {Wang}},\ }\bibfield  {title} {\bibinfo {title} {Beyond the {D}rude model:
  Surface and nonlocal effects in near-field radiative heat transfer and the
  {C}asimir puzzle},\ }\href {https://doi.org/10.1103/PhysRevB.111.245404}
  {\bibfield  {journal} {\bibinfo  {journal} {Phys. Rev. B}\ }\textbf {\bibinfo
  {volume} {111}},\ \bibinfo {pages} {245404} (\bibinfo {year}
  {2025})}\BibitemShut {NoStop}%
\bibitem [{\citenamefont {Haug}\ and\ \citenamefont
  {Jauho}(2008)}]{Haug_Jauho}%
  \BibitemOpen
  \bibfield  {author} {\bibinfo {author} {\bibfnamefont {H.}~\bibnamefont
  {Haug}}\ and\ \bibinfo {author} {\bibfnamefont {A.-P.}\ \bibnamefont
  {Jauho}},\ }\href@noop {} {\emph {\bibinfo {title} {Quantum kinetics in
  transport and optics of semiconductors}}},\ Vol.~\bibinfo {volume} {2}\
  (\bibinfo  {publisher} {Springer},\ \bibinfo {year} {2008})\BibitemShut
  {NoStop}%
\bibitem [{\citenamefont {Palik}\ \emph {et~al.}(1976)\citenamefont {Palik},
  \citenamefont {Kaplan}, \citenamefont {Gammon}, \citenamefont {Kaplan},
  \citenamefont {Wallis},\ and\ \citenamefont {Quinn}}]{InSb_76}%
  \BibitemOpen
  \bibfield  {author} {\bibinfo {author} {\bibfnamefont {E.~D.}\ \bibnamefont
  {Palik}}, \bibinfo {author} {\bibfnamefont {R.}~\bibnamefont {Kaplan}},
  \bibinfo {author} {\bibfnamefont {R.~W.}\ \bibnamefont {Gammon}}, \bibinfo
  {author} {\bibfnamefont {H.}~\bibnamefont {Kaplan}}, \bibinfo {author}
  {\bibfnamefont {R.~F.}\ \bibnamefont {Wallis}},\ and\ \bibinfo {author}
  {\bibfnamefont {J.~J.}\ \bibnamefont {Quinn}},\ }\bibfield  {title} {\bibinfo
  {title} {Coupled surface magnetoplasmon-optic-phonon polariton modes on
  {I}n{S}b},\ }\href {https://doi.org/10.1103/PhysRevB.13.2497} {\bibfield
  {journal} {\bibinfo  {journal} {Phys. Rev. B}\ }\textbf {\bibinfo {volume}
  {13}},\ \bibinfo {pages} {2497} (\bibinfo {year} {1976})}\BibitemShut
  {NoStop}%
\bibitem [{\citenamefont {Hofmann}\ and\ \citenamefont
  {Das~Sarma}(2016)}]{WSM_SPP16}%
  \BibitemOpen
  \bibfield  {author} {\bibinfo {author} {\bibfnamefont {J.}~\bibnamefont
  {Hofmann}}\ and\ \bibinfo {author} {\bibfnamefont {S.}~\bibnamefont
  {Das~Sarma}},\ }\bibfield  {title} {\bibinfo {title} {Surface plasmon
  polaritons in topological {W}eyl semimetals},\ }\href
  {https://doi.org/10.1103/PhysRevB.93.241402} {\bibfield  {journal} {\bibinfo
  {journal} {Phys. Rev. B}\ }\textbf {\bibinfo {volume} {93}},\ \bibinfo
  {pages} {241402(R)} (\bibinfo {year} {2016})}\BibitemShut {NoStop}%
\bibitem [{\citenamefont {Kotov}\ and\ \citenamefont
  {Lozovik}(2018)}]{Kotov18}%
  \BibitemOpen
  \bibfield  {author} {\bibinfo {author} {\bibfnamefont {O.~V.}\ \bibnamefont
  {Kotov}}\ and\ \bibinfo {author} {\bibfnamefont {Y.~E.}\ \bibnamefont
  {Lozovik}},\ }\bibfield  {title} {\bibinfo {title} {Giant tunable
  nonreciprocity of light in {W}eyl semimetals},\ }\href
  {https://doi.org/10.1103/PhysRevB.98.195446} {\bibfield  {journal} {\bibinfo
  {journal} {Phys. Rev. B}\ }\textbf {\bibinfo {volume} {98}},\ \bibinfo
  {pages} {195446} (\bibinfo {year} {2018})}\BibitemShut {NoStop}%
\bibitem [{\citenamefont {Li}\ \emph {et~al.}(2020)\citenamefont {Li},
  \citenamefont {Koo}, \citenamefont {Ning}, \citenamefont {Li}, \citenamefont
  {Miao}, \citenamefont {Min}, \citenamefont {Zhu}, \citenamefont {Wang},
  \citenamefont {Alem}, \citenamefont {Liu}, \citenamefont {Mao},\ and\
  \citenamefont {Yan}}]{WSM_AHE_20nc}%
  \BibitemOpen
  \bibfield  {author} {\bibinfo {author} {\bibfnamefont {P.}~\bibnamefont
  {Li}}, \bibinfo {author} {\bibfnamefont {J.}~\bibnamefont {Koo}}, \bibinfo
  {author} {\bibfnamefont {W.}~\bibnamefont {Ning}}, \bibinfo {author}
  {\bibfnamefont {J.}~\bibnamefont {Li}}, \bibinfo {author} {\bibfnamefont
  {L.}~\bibnamefont {Miao}}, \bibinfo {author} {\bibfnamefont {L.}~\bibnamefont
  {Min}}, \bibinfo {author} {\bibfnamefont {Y.}~\bibnamefont {Zhu}}, \bibinfo
  {author} {\bibfnamefont {Y.}~\bibnamefont {Wang}}, \bibinfo {author}
  {\bibfnamefont {N.}~\bibnamefont {Alem}}, \bibinfo {author} {\bibfnamefont
  {C.-X.}\ \bibnamefont {Liu}}, \bibinfo {author} {\bibfnamefont
  {Z.}~\bibnamefont {Mao}},\ and\ \bibinfo {author} {\bibfnamefont
  {B.}~\bibnamefont {Yan}},\ }\bibfield  {title} {\bibinfo {title} {Giant room
  temperature anomalous {H}all effect and tunable topology in a ferromagnetic
  topological semimetal {C}o$_2${M}n{A}l},\ }\href
  {https://doi.org/10.1038/s41467-020-17174-9} {\bibfield  {journal} {\bibinfo
  {journal} {Nat. Commun.}\ }\textbf {\bibinfo {volume} {11}},\ \bibinfo
  {pages} {3476} (\bibinfo {year} {2020})}\BibitemShut {NoStop}%
\bibitem [{\citenamefont {Tang}\ \emph
  {et~al.}(2021{\natexlab{b}})\citenamefont {Tang}, \citenamefont {Chen},\ and\
  \citenamefont {Zhang}}]{GT_WSM}%
  \BibitemOpen
  \bibfield  {author} {\bibinfo {author} {\bibfnamefont {G.}~\bibnamefont
  {Tang}}, \bibinfo {author} {\bibfnamefont {J.}~\bibnamefont {Chen}},\ and\
  \bibinfo {author} {\bibfnamefont {L.}~\bibnamefont {Zhang}},\ }\bibfield
  {title} {\bibinfo {title} {Twist-induced control of near-field heat radiation
  between magnetic {W}eyl semimetals},\ }\href
  {https://doi.org/10.1021/acsphotonics.0c01945} {\bibfield  {journal}
  {\bibinfo  {journal} {ACS Photonics}\ }\textbf {\bibinfo {volume} {8}},\
  \bibinfo {pages} {443} (\bibinfo {year} {2021}{\natexlab{b}})}\BibitemShut
  {NoStop}%
\bibitem [{SM()}]{SM}%
  \BibitemOpen
  \href@noop {} {\bibinfo {title} {{See Supplemental Material at [URL] for
  details in calculating the photonic Green’s function which includes
  Refs.~\cite{Sipe87, Eckhardt84, Kruger12}.}}}\BibitemShut {Stop}%
\bibitem [{\citenamefont {Shapiro}(2017)}]{lateral_17}%
  \BibitemOpen
  \bibfield  {author} {\bibinfo {author} {\bibfnamefont {B.}~\bibnamefont
  {Shapiro}},\ }\bibfield  {title} {\bibinfo {title} {Fluctuation-induced
  forces in the presence of mobile carrier drift},\ }\href
  {https://doi.org/10.1103/PhysRevB.96.075407} {\bibfield  {journal} {\bibinfo
  {journal} {Phys. Rev. B}\ }\textbf {\bibinfo {volume} {96}},\ \bibinfo
  {pages} {075407} (\bibinfo {year} {2017})}\BibitemShut {NoStop}%
\bibitem [{\citenamefont {Gelbwaser-Klimovsky}\ \emph
  {et~al.}(2021)\citenamefont {Gelbwaser-Klimovsky}, \citenamefont {Graham},
  \citenamefont {Kardar},\ and\ \citenamefont {Kr\"uger}}]{lateral_21}%
  \BibitemOpen
  \bibfield  {author} {\bibinfo {author} {\bibfnamefont {D.}~\bibnamefont
  {Gelbwaser-Klimovsky}}, \bibinfo {author} {\bibfnamefont {N.}~\bibnamefont
  {Graham}}, \bibinfo {author} {\bibfnamefont {M.}~\bibnamefont {Kardar}},\
  and\ \bibinfo {author} {\bibfnamefont {M.}~\bibnamefont {Kr\"uger}},\
  }\bibfield  {title} {\bibinfo {title} {Near field propulsion forces from
  nonreciprocal media},\ }\href
  {https://doi.org/10.1103/PhysRevLett.126.170401} {\bibfield  {journal}
  {\bibinfo  {journal} {Phys. Rev. Lett.}\ }\textbf {\bibinfo {volume} {126}},\
  \bibinfo {pages} {170401} (\bibinfo {year} {2021})}\BibitemShut {NoStop}%
\bibitem [{\citenamefont {Zhu}\ \emph {et~al.}(2024)\citenamefont {Zhu},
  \citenamefont {Tang}, \citenamefont {Zhang},\ and\ \citenamefont
  {Chen}}]{GT24-2}%
  \BibitemOpen
  \bibfield  {author} {\bibinfo {author} {\bibfnamefont {H.}~\bibnamefont
  {Zhu}}, \bibinfo {author} {\bibfnamefont {G.}~\bibnamefont {Tang}}, \bibinfo
  {author} {\bibfnamefont {L.}~\bibnamefont {Zhang}},\ and\ \bibinfo {author}
  {\bibfnamefont {J.}~\bibnamefont {Chen}},\ }\bibfield  {title} {\bibinfo
  {title} {Current-induced near-field radiative energy, linear-momentum, and
  angular-momentum transfer},\ }\href
  {https://doi.org/10.1103/PhysRevB.109.075413} {\bibfield  {journal} {\bibinfo
   {journal} {Phys. Rev. B}\ }\textbf {\bibinfo {volume} {109}},\ \bibinfo
  {pages} {075413} (\bibinfo {year} {2024})}\BibitemShut {NoStop}%
\bibitem [{\citenamefont {A.~Grinberg}\ and\ \citenamefont
  {Luryi}(1988)}]{photon_drag88}%
  \BibitemOpen
  \bibfield  {author} {\bibinfo {author} {\bibfnamefont {A.}~\bibnamefont
  {A.~Grinberg}}\ and\ \bibinfo {author} {\bibfnamefont {S.}~\bibnamefont
  {Luryi}},\ }\bibfield  {title} {\bibinfo {title} {Theory of the photon-drag
  effect in a two-dimensional electron gas},\ }\href
  {https://doi.org/10.1103/PhysRevB.38.87} {\bibfield  {journal} {\bibinfo
  {journal} {Phys. Rev. B}\ }\textbf {\bibinfo {volume} {38}},\ \bibinfo
  {pages} {87} (\bibinfo {year} {1988})}\BibitemShut {NoStop}%
\bibitem [{\citenamefont {Strait}\ \emph {et~al.}(2019)\citenamefont {Strait},
  \citenamefont {Holland}, \citenamefont {Zhu}, \citenamefont {Zhang},
  \citenamefont {Ilic}, \citenamefont {Agrawal}, \citenamefont {Pacifici},\
  and\ \citenamefont {Lezec}}]{photon_drag19}%
  \BibitemOpen
  \bibfield  {author} {\bibinfo {author} {\bibfnamefont {J.~H.}\ \bibnamefont
  {Strait}}, \bibinfo {author} {\bibfnamefont {G.}~\bibnamefont {Holland}},
  \bibinfo {author} {\bibfnamefont {W.}~\bibnamefont {Zhu}}, \bibinfo {author}
  {\bibfnamefont {C.}~\bibnamefont {Zhang}}, \bibinfo {author} {\bibfnamefont
  {B.~R.}\ \bibnamefont {Ilic}}, \bibinfo {author} {\bibfnamefont
  {A.}~\bibnamefont {Agrawal}}, \bibinfo {author} {\bibfnamefont
  {D.}~\bibnamefont {Pacifici}},\ and\ \bibinfo {author} {\bibfnamefont
  {H.~J.}\ \bibnamefont {Lezec}},\ }\bibfield  {title} {\bibinfo {title}
  {Revisiting the photon-drag effect in metal films},\ }\href
  {https://doi.org/10.1103/PhysRevLett.123.053903} {\bibfield  {journal}
  {\bibinfo  {journal} {Phys. Rev. Lett.}\ }\textbf {\bibinfo {volume} {123}},\
  \bibinfo {pages} {053903} (\bibinfo {year} {2019})}\BibitemShut {NoStop}%
\bibitem [{\citenamefont {Mironov}\ \emph {et~al.}(2024)\citenamefont
  {Mironov}, \citenamefont {Mel'nikov},\ and\ \citenamefont
  {Buzdin}}]{photon_drag24}%
  \BibitemOpen
  \bibfield  {author} {\bibinfo {author} {\bibfnamefont {S.~V.}\ \bibnamefont
  {Mironov}}, \bibinfo {author} {\bibfnamefont {A.~S.}\ \bibnamefont
  {Mel'nikov}},\ and\ \bibinfo {author} {\bibfnamefont {A.~I.}\ \bibnamefont
  {Buzdin}},\ }\bibfield  {title} {\bibinfo {title} {ac {H}all effect and
  photon drag of superconducting condensates},\ }\href
  {https://doi.org/10.1103/PhysRevLett.132.096001} {\bibfield  {journal}
  {\bibinfo  {journal} {Phys. Rev. Lett.}\ }\textbf {\bibinfo {volume} {132}},\
  \bibinfo {pages} {096001} (\bibinfo {year} {2024})}\BibitemShut {NoStop}%
\bibitem [{\citenamefont {Moiseenko}\ \emph {et~al.}(2025)\citenamefont
  {Moiseenko}, \citenamefont {Svintsov},\ and\ \citenamefont
  {Devizorova}}]{photon_drag25}%
  \BibitemOpen
  \bibfield  {author} {\bibinfo {author} {\bibfnamefont {I.~M.}\ \bibnamefont
  {Moiseenko}}, \bibinfo {author} {\bibfnamefont {D.~A.}\ \bibnamefont
  {Svintsov}},\ and\ \bibinfo {author} {\bibfnamefont {Z.~A.}\ \bibnamefont
  {Devizorova}},\ }\bibfield  {title} {\bibinfo {title} {Electromagnetic drag
  in partly gated 2{D} electron system via highly confined screened plasmons},\
  }\href {https://doi.org/10.1134/S0021364025609480} {\bibfield  {journal}
  {\bibinfo  {journal} {JETP Lett.}\ }\textbf {\bibinfo {volume} {122}},\
  \bibinfo {pages} {743} (\bibinfo {year} {2025})}\BibitemShut {NoStop}%
\bibitem [{\citenamefont {Dehaghi}\ and\ \citenamefont
  {Zhu}(2025)}]{photon_Nernst25}%
  \BibitemOpen
  \bibfield  {author} {\bibinfo {author} {\bibfnamefont {A.~K.}\ \bibnamefont
  {Dehaghi}}\ and\ \bibinfo {author} {\bibfnamefont {L.}~\bibnamefont {Zhu}},\
  }\href@noop {} {\bibinfo {title} {Near-field photon {N}ernst effect}}
  (\bibinfo {year} {2025}),\ \Eprint {https://arxiv.org/abs/arXiv:2511.20372}
  {arXiv:2511.20372} \BibitemShut {NoStop}%
\bibitem [{\citenamefont {Sipe}(1987)}]{Sipe87}%
  \BibitemOpen
  \bibfield  {author} {\bibinfo {author} {\bibfnamefont {J.~E.}\ \bibnamefont
  {Sipe}},\ }\bibfield  {title} {\bibinfo {title} {New {G}reen-function
  formalism for surface optics},\ }\href
  {https://doi.org/10.1364/JOSAB.4.000481} {\bibfield  {journal} {\bibinfo
  {journal} {J. Opt. Soc. Am. B}\ }\textbf {\bibinfo {volume} {4}},\ \bibinfo
  {pages} {481} (\bibinfo {year} {1987})}\BibitemShut {NoStop}%
\bibitem [{\citenamefont {Eckhardt}(1984)}]{Eckhardt84}%
  \BibitemOpen
  \bibfield  {author} {\bibinfo {author} {\bibfnamefont {W.}~\bibnamefont
  {Eckhardt}},\ }\bibfield  {title} {\bibinfo {title} {Macroscopic theory of
  electromagnetic fluctuations and stationary radiative heat transfer},\ }\href
  {https://doi.org/10.1103/PhysRevA.29.1991} {\bibfield  {journal} {\bibinfo
  {journal} {Phys. Rev. A}\ }\textbf {\bibinfo {volume} {29}},\ \bibinfo
  {pages} {1991} (\bibinfo {year} {1984})}\BibitemShut {NoStop}%
\bibitem [{\citenamefont {Kr\"uger}\ \emph {et~al.}(2012)\citenamefont
  {Kr\"uger}, \citenamefont {Bimonte}, \citenamefont {Emig},\ and\
  \citenamefont {Kardar}}]{Kruger12}%
  \BibitemOpen
  \bibfield  {author} {\bibinfo {author} {\bibfnamefont {M.}~\bibnamefont
  {Kr\"uger}}, \bibinfo {author} {\bibfnamefont {G.}~\bibnamefont {Bimonte}},
  \bibinfo {author} {\bibfnamefont {T.}~\bibnamefont {Emig}},\ and\ \bibinfo
  {author} {\bibfnamefont {M.}~\bibnamefont {Kardar}},\ }\bibfield  {title}
  {\bibinfo {title} {Trace formulas for nonequilibrium casimir interactions,
  heat radiation, and heat transfer for arbitrary objects},\ }\href
  {https://doi.org/10.1103/PhysRevB.86.115423} {\bibfield  {journal} {\bibinfo
  {journal} {Phys. Rev. B}\ }\textbf {\bibinfo {volume} {86}},\ \bibinfo
  {pages} {115423} (\bibinfo {year} {2012})}\BibitemShut {NoStop}%
\bibitem [{\citenamefont {Fradkin}(2021)}]{QFT-Fradkin}%
  \BibitemOpen
  \bibfield  {author} {\bibinfo {author} {\bibfnamefont {E.}~\bibnamefont
  {Fradkin}},\ }\href@noop {} {\emph {\bibinfo {title} {Quantum field theory:
  an integrated approach}}}\ (\bibinfo  {publisher} {Princeton University
  Press},\ \bibinfo {year} {2021})\BibitemShut {NoStop}%
\bibitem [{\citenamefont {Lifshitz}\ and\ \citenamefont
  {Pitaevskii}(2013)}]{Lifshitz_book}%
  \BibitemOpen
  \bibfield  {author} {\bibinfo {author} {\bibfnamefont {E.~M.}\ \bibnamefont
  {Lifshitz}}\ and\ \bibinfo {author} {\bibfnamefont {L.~P.}\ \bibnamefont
  {Pitaevskii}},\ }\href@noop {} {\emph {\bibinfo {title} {Statistical Physics:
  Theory of the Condensed State}}},\ Vol.~\bibinfo {volume} {9}\ (\bibinfo
  {publisher} {Elsevier},\ \bibinfo {year} {2013})\BibitemShut {NoStop}%
\end{thebibliography}

\section*{End Matter}
The total Hamiltonian of the system is partitioned as
\begin{equation}
  H_{\rm tot} = H_{\rm ph} + H_1 + H_{1, \rm ph} .
\end{equation}
We adopt the temporal gauge by setting the scalar potential to zero so that the
electromagnetic Hamiltonian is expressed in terms of the vector potential $\bm{A}$
as~\cite{QFT-Fradkin, Lifshitz_book, JSW23}
\begin{equation} \label{H_ph}
  H_{\rm ph} = \frac{1}{2} \int d^3\bm{r} \left[ \epsilon_0 \epsilon_r \big(
  \partial_t \bm{A} \big)^2 + \mu_0^{-1} \big( \nabla \times \bm{A} \big)^2 \right] ,
\end{equation}
with $\epsilon_0$ and $\mu_0$ the vacuum permittivity and permeability. The
space-dependent relative permittivity is $\epsilon_r = \epsilon_{\rm MO}$ for $z<0$
which is filled with magneto-optic medium and $\epsilon_r = 1$ for $z>0$.
The two-dimensional metal sheet is described using a tight-binding model with
\begin{equation} 
  H_1 = \sum_{i, j} c_i^\dag H_{ij} c_j ,
\end{equation}
where $i$ and $j$ label the electron sites at positions $\bm{r}_i$ and $\bm{r}_j$. 
The electron-photon interaction for the metal sheet is considered in $H_{1, \rm ph}$.
Gauge invariance is ensured by the Peierls substitution, which couples the electrons to
the photonic field with~\cite{JSW23}
\begin{equation} \label{Peierls}
  H_1 + H_{1, \rm ph} = \sum_{i, j} c_i^\dag  H_{ij} c_j
  \exp\left[ \frac{e_0}{i\hbar} \int_{\bm{r}_j}^{{\bm{r}_i}} d\bm{r} \cdot \bm{A}(\bm{r})
  \right] .
\end{equation}
To get the explicit interaction Hamiltonian, the line integral is approximated with the
trapezoidal rule
\begin{equation}
  \int_{\bm{r}_j}^{{\bm{r}_i}} d\bm{r} \cdot \bm{A}(\bm{r}) 
  \approx \frac{1}{2} (\bm{A}_i + \bm{A}_j) \cdot (\bm{r}_i - \bm{r}_j) ,
\end{equation}
with the notation $\bm{A}_i \equiv \bm{A}(\bm{r}_i)$. 
Performing Taylor expansion to the exponential in Eq.~\eqref{Peierls} up to the second
order in the lattice constant, the interaction Hamiltonian is written as
\begin{equation} \label{H_int} 
  H_{1,\rm ph} = H_{\rm para} + H_{\rm dia} ,
\end{equation}
which consists of paramagnetic and diamagnetic contributions. 
The paramagnetic contribution is given explicitly in real space as~\cite{JSW23}
\begin{equation} 
  H_{\rm para} = \frac{e_0}{2} \sum_{i,j, l,\mu} c_i^\dag v_{ij}^\mu (\delta_{il} +
  \delta_{jl}) A_{l \mu} c_j ,
\end{equation}
where the indices $i$, $j$, and $l$ denote the electron sites, and $\mu =x,y$ labels the
Cartesian directions. Transforming to the wavevector representation, the electronic
Hamiltonian becomes
\begin{equation}
  H_1 = \sum_{\bm{k}} \varepsilon_{\bm{k}} c_{\bm{k}}^{\dag}c_{\bm{k}} ,
\end{equation}
where $\varepsilon_{\bm{k}}$ is the band dispersion. The paramagnetic interaction is expressed
as
\begin{equation}
  H_{\rm para} = -e_0 \sum_{\bm{q}, \bm{k}} \bm{v}(\bm{k},\bm{q}) \cdot \bm{A}(\bm{q})
  c_{\bm{k}+\bm{q}}^{\dag}c_{\bm{k}} ,
\end{equation}
with the symmetrized velocity vertex as
\begin{equation}
  \bm{v}(\bm{k},\bm{q}) = \big[ \bm{v}(\bm{k}) + \bm{v}(\bm{k} + \bm{q}) \big] / 2 .
\end{equation}

In the absence of the electron-photon interaction, the non-interacting retarded and
advanced electronic Green's functions are given by
\begin{equation}
  g^r(\bm{k},E) = \big[ g^a(\bm{k},E)\big]^\dag
  = \big( E - \varepsilon_{\bm{k}} + i\eta \big)^{-1} ,
\end{equation}
where $\hbar/\eta$ is the electron relaxation time in the metal sheet. 
The Keldysh equations for the non-interacting lesser and greater electronic Green's
functions are
\begin{equation} \label{g<>}
  g^{<,>}(\bm{k},E) = g^r(\bm{k},E) \Sigma_B^{<,>}(E) g^a(\bm{k},E) ,
\end{equation}
with the bath self-energies given by
\begin{equation}
  \Sigma_B^<(E) = 2i\eta f(E), \quad \Sigma_B^>(E) = 2i\eta \big[f(E) -1 \big] ,
\end{equation}
where $f(E)$ is the Fermi-Dirac distribution function given by Eq.~\eqref{Fermi-Dirac}. 
Including the electron-photon interaction, the retarded Green's function obey the Dyson
equation 
\begin{equation}
  G^r(\bm{k},E) = g^r(\bm{k},E) + g^r(\bm{k},E) \Sigma^r_F(\bm{k},E) G^r(\bm{k},E) ,
\end{equation}
where the Fock-like self-energy $\Sigma_F$ encodes the coupling with photons. 
The advanced Green's function $G^a$ follows from Hermitian conjugation: 
$G^a(\bm{k},E) = \big[G^r(\bm{k},E)\big]^\dag$.
The lesser Green's function which describes the electron distribution follows the Keldysh
equation
\begin{equation} \label{G<}
  G^<(\bm{k},E) = G^r(\bm{k},E) \big[ \Sigma_B^<(\bm{k},E) + \Sigma_F^<(\bm{k},E) \big]
  G^a(\bm{k},E) .
\end{equation}
The Fock self-energy, which accounts for the paramagnetic interaction, is given by the
convolution of the non-interacting electronic Green's functions with the photonic Green's
functions. The explicit expressions for its retarded and lesser components are,
respectively,
\begin{align}
  \Sigma^r_F(\bm{k},E) = & \int d\Gamma \sum_{\mu\nu} v_\mu(\bm{k},\bm{q})
  v_\nu(\bm{k},\bm{q}) \notag \\
  \big[ & g^r(\bm{k}+\bm{q},E+\hbar\omega) D_{1, \mu\nu}^<(-\bm{q},-\omega)  \notag \\
  + & g^>(\bm{k}+\bm{q},E+\hbar\omega) D_{1, \mu\nu}^r(-\bm{q},-\omega) \big]
  \label{Sigma_r} ,
\end{align}
and
\begin{align}
  \Sigma^<_F(\bm{k},E) = & \int d\Gamma \sum_{\mu\nu} v_\mu(\bm{k},\bm{q})
  v_\nu(\bm{k},\bm{q}) \notag \\
  & g^<(\bm{k}+\bm{q},E+\hbar\omega) D_{1, \mu\nu}^<(-\bm{q},-\omega) ,
\end{align}
where $\mu,\nu =x,y$ label the Cartesian directions and the integration measure is
\begin{equation}
  \int d\Gamma = i\hbar e_0^2 \int_{-\infty}^\infty \frac{d\omega}{2\pi} \int
  \frac{d^2\bm{q}}{(2\pi)^2} .
\end{equation}
The photonic Green's function is obtained as (See the Supplemental Material~\cite{SM} for
details)
\begin{equation} \label{D1<_near}
  D_1^< = D_{\rm eq}^< + D_{\rm neq}^< ,
\end{equation}
with the equilibrium part $D_{\rm eq}^< = (D_1^r - D_1^a) N_1$ and the nonequilibrium
part
\begin{equation} 
  D_{\rm neq}^<(\bm{q}, \omega) = -\frac{i|\gamma_0|}{\epsilon_0\omega^2} \Phi(\bm{q},
  \omega) N_{21}(\omega) \big( \hat{\bm{q}}^T \hat{\bm{q}} \big) .
\end{equation}

The electric current along $x$ direction is expressed in terms of the lesser electronic
Green's function as
\begin{equation}
  I_e = 4 e_0 i \int_{-\infty}^{\infty} \frac{dE}{2\pi} 
  \int \frac{d^2\bm{k}}{(2\pi)^2} v_x({\bm{k}}) G^<(\bm{k}, E) ,
\end{equation}
where the factor of $4$ is included to account for the spin and valley degeneracies. 
By keeping both the Fock self-energy and damping $\eta$ to the first order, the electric
current can be expressed as
\begin{align}
  I_e = -4 & \hbar e_0^3 
   \int_{-\infty}^{\infty} \frac{dE}{2\pi} \int \frac{d^2\bm{k}}{(2\pi)^2}v_x({\bm{k}}) 
  \int_{q>k_0} \frac{d^2\bm{q}}{(2\pi)^2} \int_{-\infty}^\infty \frac{d\omega}{2\pi}
  \notag \\
   & \Big\{ 2i\ {\rm Im}\big[
   g^r(\bm{k}, E) g^r(\bm{k}+\bm{q},E+\hbar\omega) g^<(\bm{k}, E) \big]  \notag \\
  &\qquad \
  + g^r(\bm{k}, E) g^<(\bm{k}+\bm{q},E+\hbar\omega) g^a(\bm{k}, E) \Big\} \notag \\
  &\quad \sum_{\mu\nu} v_\mu(\bm{k},\bm{q}) v_\nu(\bm{k},\bm{q})
   D_{{\rm neq}, \mu\nu}^<(-\bm{q},-\omega)  .
\end{align}
Here, the contribution from the third line of Eq.~\eqref{Sigma_r} has been neglected, as
it is of order $\eta^2$. 
The contribution from $D_{{\rm eq}, \mu\nu}^<$ vanishes due to the symmetry 
$D_{\mu\nu}^r(\bm{q}, \omega) = D_{\nu\mu}^a(-\bm{q}, -\omega)$.
It is noted that
\begin{align}
  & \sum_{\mu\nu} v_\mu(\bm{k},\bm{q}) v_\nu(\bm{k},\bm{q})
  D_{{\rm neq}, \mu\nu}^<(-\bm{q},-\omega) \notag \\
  =-& \frac{i |\gamma_0|}{\epsilon_0 \omega^2} \Phi(\bm{q}, \omega) 
  \left[ \bm{v}(\bm{k},\bm{q}) \cdot \hat{\bm{q}} \right]^2 N_{21}(\omega) ,
\end{align}
where we have used the symmetries of the photon flux factor $\Phi$ and the Bose distribution difference $N_{21}$:
\begin{equation}\label{symmetries}
  \Phi(-\bm{q},-\omega) = -\Phi(\bm{q},\omega), \quad N_{21}(-\omega) = - N_{21}(\omega).  
\end{equation}
Applying the residue theorem to integrate over the electron energy $E$, the electric
current reduces to
\begin{align}
  I_e = -\frac{4 e_0}{\eta} & \int \frac{d^2\bm{k}}{(2\pi)^2} v_x({\bm{k}}) 
  \int_{q>k_0} \frac{d^2\bm{q}}{(2\pi)^2} \int_{-\infty}^\infty \frac{d\omega}{2\pi}
  \notag \\
  & L(\bm{k},\bm{q}, \omega) M(\bm{k},\bm{q}, \omega) N_{21}(\omega) ,
\end{align}
where the electronic transition factor $L$  and the photon interaction strength $M$ are given by
Eqs.~\eqref{L} and \eqref{M}, respectively. Using again Eq.~\eqref{symmetries},
this expression simplifies further to Eq.~\eqref{Ie}.

\end{document}